# Information Cost Tradeoffs for Augmented Index and Streaming Language Recognition

Amit Chakrabarti[*]     Graham Cormode[†]     Ranganath Kondapally[‡]     Andrew McGregor[§]

**Abstract**

This paper makes three main contributions to the theory of communication complexity and stream computation. First, we present new bounds on the information complexity of AUGMENTED-INDEX. In contrast to analogous results for INDEX by Jain, Radhakrishnan and Sen [*J. ACM*, 2009], we have to overcome the significant technical challenge that protocols for AUGMENTED-INDEX may violate the "rectangle property" due to the inherent input sharing. Second, we use these bounds to resolve an open problem of Magniez, Mathieu and Nayak [*STOC*, 2010] that asked about the multi-pass complexity of recognizing Dyck languages. This results in a natural separation between the standard multi-pass model and the multi-pass model that permits reverse passes. Third, we present the first *passive memory checkers* that verify the interaction transcripts of priority queues, stacks, and double-ended queues. We obtain tight upper and lower bounds for these problems, thereby addressing an important sub-class of the memory checking framework of Blum et al. [*Algorithmica*, 1994].

[*]Dartmouth College. ac@cs.dartmouth.edu. Work supported by NSF Grant IIS-0916565 and by Dartmouth College startup funds.

[†]AT&T Labs–Research. graham@research.att.com

[‡]Dartmouth College. rangak@cs.dartmouth.edu. Work supported by NSF Grant IIS-0916565 and by Dartmouth College startup funds.

[§]University of Massachusetts, Amherst. mcgregor@cs.umass.edu. Work supported by NSF CAREER Award CCF-0953754 and by University of Massachusetts startup funds.

# 1 Introduction

In recent work, Magniez, Mathieu and Nayak [15] considered the streaming complexity of language recognition. That is, given a string $\sigma$ of length $n$, what is the (randomized) space complexity of a recognizer for a language $L$ that is allowed only sequential access to $\sigma$? This question can be viewed as a generalization of the classic notion of regularity of languages: one now considers automata that are allowed (1) randomization, and (2) a variable number of states that may depend on the input length. Their main result provided near-matching bounds for single-pass recognizers for DYCK(2), the language of properly nested parentheses of two kinds. In this paper, we look at the broader question and present the first *multi-pass* space lower bounds for several languages, including DYCK(2), resolving an open question of theirs. We also study the complexity of languages that arise in the context of *memory checking* [5], and present tight upper and lower bounds for them. Our key technical contributions rely on a new understanding of the information complexity of the augmented index problem, which leads to these multi-pass lower bounds.

**Background, Augmented Index, and a New Lower Bound.** The INDEX problem is one of a handful of fundamental problems in communication complexity [14]: Alice has a string $x \in \{0,1\}^n$, and Bob has an index $k \in [n]$; the players wish to determine the $k$th bit of $x$, written as $x_k$. It is easy to show that the problem is "hard" — requiring $\Omega(n)$ communication — when messages only go from Alice to Bob, and is "easy" — solvable using $O(\log n)$ communication — without this restriction. The lower bound extends to randomized constant-error protocols [1]. This makes INDEX the canonical hard-for-one-way, easy-for-two-way communication problem. Is there really anything new to say about such a fundamental problem?

As it turns out, there is, provided one asks the right questions. Since INDEX is an asymmetric problem, it makes sense to ask for the best possible tradeoff between the number, $a$, of bits communicated by Alice, and the number, $b$, communicated by Bob. As shown by Miltersen et al. [16], we must have $a \geq n/2^{O(b)}$, and a simple two-round Bob $\to$ Alice $\to$ Bob protocol (with Bob announcing the output) shows that $a \leq \lceil n/2^b \rceil$ is achievable. A more nuanced question asks for the best tradeoff of *information revealed* by each player to the other in a protocol for INDEX, also called the *information costs* of Alice and Bob (we shall soon formally define these). In principle, this tradeoff could have been better, as it is possible for messages to reveal less information than their length. This issue was considered (in a more general quantum communication setting) by Jain, Radhakrishnan and Sen [12] who called this the "privacy tradeoff" for the problem, and showed that $a \geq n/2^{O(b)}$ still holds, where $a$ and $b$ now represent information costs.

Such an information cost tradeoff opens up interesting possibilities for applications to lower bounds for more complex problems, via the direct sum properties of this measure [4, 6]. One such application is the aforementioned DYCK(2) lower bound. However, the tradeoff theorem of Jain et al. is not strong enough to obtain the required direct sum result. One needs a tradeoff lower bound in a variant of INDEX where Alice and Bob have much more "help," in two ways. First, we relax INDEX so that Bob additionally gets to see the length-$(k-1)$ prefix of Alice's input; the resulting variant has been called AUGMENTED-INDEX [10, 13] and the one-way communication lower bound easily extends to it [3]. Second, in our variant of AUGMENTED-INDEX, Bob also gets a *check bit* $c \in \{0,1\}$ and must verify that $x_k = c$. This second twist clearly does not matter when considering communication complexity, but for us it makes a huge difference, because our applications require that we measure information cost under an "easy" distribution, where $x_k$ always equals $c$.

With this background, we state our main theorem informally. A formal version appears as Theorem 2.3, after the necessary definitions.

**Theorem 1.1** (Informal)**.** *In a randomized communication protocol for* AUGMENTED-INDEX *with two-sided error at most* $1/\log^2 n$, *either Alice reveals* $\Omega(n)$ *information about her input, or Bob reveals* $\Omega(1)$ *information about his, where information is measured according to an "easy" input distribution.*

The natural point of comparison is a similar theorem of Magniez et al. [15], that works only for restricted protocols: Alice must be deterministic (thus, her information cost is just her usual communication cost), and the protocol must be two-round with an Alice $\to$ Bob $\to$ Alice communication pattern. Under these conditions, for errors below $O(1/n^2)$, they show that either Alice sends $\Omega(n)$ bits, or else Bob reveals $\Omega(\log n)$ information. Notice that this theorem is not quite a special case of ours, because of the higher lower bound on Bob's information cost. This is inevitable: for a general communication pattern, one cannot obtain a tradeoff that strong, because of the aforementioned $a \le \lceil n/2^b \rceil$ upper bound. However, we suspect that the optimum tradeoff lower bound is in fact of the form $a \ge n/2^{\widetilde{O}(b)}$, where the $\widetilde{O}$-notation hides factors polylogarithmic in $b$, and we leave this conjectured generalization of our theorem as an open problem.

**Ramifications: Streaming Language Recognition.** In the streaming model, we have one-way access to input and working memory sublinear in the input size $N$. Historically, the problems considered in it have focused on estimating statistics. Recognizing structural properties of strings is just as natural a problem in this model, and yet such language recognition problems have only recently been considered. It transpires that AUGMENTED-INDEX has a key role to play in proving bounds here.

A first application is direct: following [15], a two-step argument shows $\Omega(\sqrt{N})$ lower bounds for the multi-pass streaming complexity of Dyck languages. We first plug Theorem 1.1 into a direct sum theorem, which lower bounds the communication cost of a problem we call MULTI-AI (for "multiple copies of AUGMENTED-INDEX"). We then reduce MULTI-AI to, e.g., DYCK(2). The direct sum theorem is a natural extension to multiple passes of a similar single-pass theorem ([15], where the authors called the relevant problems ASCENSION and MOUNTAIN). Thus, on the lower bound side, our chief contribution is Theorem 1.1, and its most important consequence is the *multi-pass* nature of the resulting lower bounds. In particular, this demonstrates a curious phenomenon: an explicit, natural data stream problem that is fairly easy given two passes in opposite directions (Magniez et al. give an $O(\log^2 N)$-space algorithm), whereas it is exponentially harder if only multiple unidirectional passes are allowed.

A second application is to *memory checking*, whose study was initiated by Blum et al. [5] and continued by numerous groups including Ajtai [2], Chu et al. [8], Dwork et al. [11], and Naor and Rothblum [17]. The problem, as considered in this paper, is to observe a sequence of $N$ updates and queries to (an implementation of) a data structure, and to report whether or not the implementation operated correctly on the instance observed. A concrete example is to observe a transcript of operations on a priority queue: we see a sequence of insertions intermixed with items claimed to be the results of extractions, and the problem is to decide whether this is correct. Much of the previous work allowed the checker to be *invasive*, by modifying the inserted items and/or introducing additional read operations. However, when the checker is more realistically restricted to being completely *passive*, and can only observe, the problem becomes that of understanding the (streaming) complexity of *recognizing* valid transcripts. For instance, we define PQ to be the language of valid transcripts of priority queue operations that start and end with an empty queue. One can similarly define languages STACK and DEQUE (for double-ended queues). The invasive protocols of [5] typically modified the input items by attaching a "timestamp" to each inserted item, and this suggests variant languages PQ-TS, STACK-TS, and DEQUE-TS, where each extraction is augmented by the timestamp of its corresponding insertion. Though we briefly study these variant languages towards the end of this paper, we consider the languages without auxiliary information to be more natural from a theoretical point of view, and more applicable from a practical point of view.

We present new algorithms for these basic memory checking languages: we show that PQ, STACK, and DEQUE can each be recognized in $\tilde{O}(\sqrt{N})$ space and one pass. On the lower bound side, Theorem 1.1 and MULTI-AI again come into play, giving $\Omega(\sqrt{N})$ bounds for each of these problems, even allowing multiple passes over the transcript. We observe that our upper bound for PQ strengthens the $\tilde{O}(\sqrt{N})$ bound of Chu et al. [8] for PQ-TS. This strengthening is significant, for timestamps can radically simplify problems: we note that STACK-TS can be recognized in just $O(\log N)$ space, in marked contrast to STACK.

**Highlights.** Since we view Theorem 1.1 as our most important technical contribution, we first give a careful exposition of its proof, in Section 2. The main technical hurdle in this proof is dealing with the fact that Alice and Bob share some of the input, which breaks the useful "rectangle property." (This is reminiscent of number-on-the-forehead communication [7], where input sharing makes strong lower bounds rather hard to prove.) The highlight of our proof is the Fat Transcript Lemma (Lemma 2.6), with its careful interplay between a suitably weakened rectangle property (Lemma 2.5) and the information cost measure.

After a brief discussion (Section 3) of the direct sum theorem and its implications for MULTI-AI, we address language recognition, in Section 4. The highlight of this section is our algorithm for recognizing the language PQ. The ingenuity of the algorithm is that, rather than determining whether the interaction sequence is valid directly, the algorithm conceptually reorders inserts and extracts (in addition to some actual "local" reordering) in such a way that the new sequence is valid if and only if the original sequence is valid. This reordering procedure is designed such that small-space fingerprinting methods can be used to capture the state of the priority queue in a way they could not necessarily have been used for the original sequence.

## 2 Augmented Index and an Information Cost Tradeoff

Let $\text{AI} = \text{AI}_n$ (short for AUGMENTED-INDEX) denote the communication problem where Alice receives a string $x \in \{0,1\}^n$, Bob receives an index $k \in [n]$, the length-$(k-1)$ prefix of $x$, which we denote $x_{1:k-1}$, and a check bit $c \in \{0,1\}$, and the goal is to output $\text{AI}(x, k, c) := x_k \oplus c$, i.e., to output 1 iff $x_k \neq c$.

We now formalize the notion of information cost. For this, we consider the most general model of randomization in communication protocols: the parties may share a public coin, and separately, each party may have its own private coin. Let $P$ be such a randomized protocol for AI, let $\xi$ be a distribution on $\{0,1\}^n \times [n] \times \{0,1\}$ (effectively, a distribution on legal inputs to $P$) and let $(X, K, C) \sim \xi$. Let $R$ denote the public random string used by $P$, and let $T$ denote the transcript of messages sent by Alice and Bob (including the final output bit) in response to this random input $(X, K, C)$: note that, in general, $T$ depends on $X, K, C, R$ and the (unnamed) private random strings of the players. We define the *information cost* of $P$ under $\xi$ to be a pair of real numbers $(\text{icost}_\xi^A(P), \text{icost}_\xi^B(P))$ defined as follows:

$$\text{icost}_\xi^A(P) := \text{I}(T : X \mid K, C, R); \quad \text{icost}_\xi^B(P) := \text{I}(T : K, C \mid X, R). \tag{1}$$

In the above definition, the conditioning on $R$ is crucial, for otherwise it is simple to make these costs equal zero. It follows from the basics of information theory that, regardless of the choice of $\xi$, these costs are bounded from above by the number of bits communicated by Alice and Bob, respectively, in $P$. Thus, a tradeoff lower bound on information cost is a stronger statement than a similar tradeoff on numbers of bits communicated. We now turn to the choice of input distribution.

**Definition 2.1.** We let $\mu$ denote the uniform distribution on $\{0,1\}^n \times [n] \times \{0,1\}$. For $(X, K, C) \sim \mu$, we let $\mu_0 := \mu \mid (X_K = C)$. Note that $\mathbb{E}_\mu[\text{AI}(X, K, C)] = \frac{1}{2}$, whereas $\mathbb{E}_{\mu_0}[\text{AI}(X, K, C)] = 0$. Thus, intuitively, $\mu$ is a hard distribution for AI, whereas $\mu_0$ is an easy distribution.

We are now ready to state our main theorem. But first, we give a technical lemma that is useful in formalizing some averaging arguments in its proof.

**Lemma 2.2.** *Consider functions $f_1, \ldots, f_L : D \to \mathbb{R}^+$, and numbers $b_1, \ldots, b_L \in \mathbb{R}^+$, where $L > 0$ is an integer and $D$ is a finite domain. Let $Z$ be a random variable taking values in $D$. Then*

$$\forall i \in [L] \ \mathbb{E}[f_i(Z)] \leq b_i \implies \exists z \in D \ \forall i \in [L] \ f_i(z) \leq L b_i \,.$$

*Proof.* Choose $z$ to minimize $g(z) := \sum_{i: b_i > 0} f_i(z)/b_i$, and observe that $\mathbb{E}[g(Z)] \leq L$, so that $g(z) \leq L$. Now pick any $i \in [L]$. If $b_i = 0$, then clearly $f_i(z) = 0$. Else, $f_i(z)/b_i \leq g(z) \leq L$. □

**Theorem 2.3** (Main Theorem; formal version of Theorem 1.1). *If $P$ is a randomized protocol for $\mathrm{AI}_n$ with error at most $1/\log^2 n$ under $\mu$, then either $\mathrm{icost}^A_{\mu_0}(P) = \Omega(n)$ or $\mathrm{icost}^B_{\mu_0}(P) = \Omega(1)$. In particular, the same tradeoff holds if $P$ has worst case two-sided error at most $1/\log^2 n$.*

*Proof.* We split this proof into two parts. First, assuming the contrary, we zoom in on a specific setting of the public random string of $P$ and a single transcript that has certain "fatness" properties that play a role analogous to the "large rectangles" seen in elementary communication complexity. This part of the proof is reminiscent of arguments in Pătraşcu's proof of the lopsided set disjointness lower bound [18]. Next, and more interestingly, we use these fatness properties to derive a contradiction, in Lemma 2.6. Throughout the proof, and the rest of this section, we tacitly assume that $n$ is large enough.

Assume, to the contrary, that for every choice of constants $\delta_1$ and $\delta_2$, there exists a $(1/\log^2 n)$-error protocol $P^*$ for $\mathrm{AI}$ with $\mathrm{icost}^A_{\mu_0}(P^*) \leq \delta_1 n$ and $\mathrm{icost}^B_{\mu_0}(P^*) \leq \delta_2$. To write these conditions formally, let $T^*$ denote the transcript of $P^*$ (which uses a public random string $R$) on input $(X, K, C) \sim \mu$; we will condition on $X_K = C$ when necessary, to effectively change the input distribution to $\mu_0$. We adopt the convention that a transcript, $t$, also specifies its final output bit, $\mathrm{out}(t)$. We then have

$$\Pr[\mathrm{out}(T^*) \neq \mathrm{AI}(X, K, C)] \leq 1/\log^2 n,$$
$$\mathrm{I}(T^* : X \mid K, C, R, X_K = C) \leq \delta_1 n, \text{ and}$$
$$\mathrm{I}(T^* : K, C \mid X, R, X_K = C) \leq \delta_2.$$

These three inequalities can be interpreted as bounding the expectations of three non-negative functions of the random string $R$. Any particular setting of $R$ reduces $P^*$ to a private-coin protocol. Thus, applying Lemma 2.2 to these three inequalities, we see that there exists a *private-coin* protocol $P$ for $\mathrm{AI}$ such that, if $T$ denotes the transcript of $P$ on input $(X, K, C) \sim \mu$, then

$$\Pr[\mathrm{out}(T) \neq \mathrm{AI}(X, K, C)] \leq 3/\log^2 n, \tag{2}$$
$$\mathrm{I}(T : X \mid K, C, X_K = C) \leq 3\delta_1 n, \text{ and} \tag{3}$$
$$\mathrm{I}(T : K, C \mid X, X_K = C) \leq 3\delta_2. \tag{4}$$

Notice that $\mathrm{H}(X \mid K, C, X_K = C) = n - 1$ and $\mathrm{H}(K, C \mid X, X_K = C) = \log n$. Thus, by the characterization of mutual information in terms of entropy, we can rewrite (3) and (4) as

$$n - 1 - \mathrm{H}(X \mid T, K, C, X_K = C) \leq 3\delta_1 n, \text{ and} \tag{5}$$
$$\log n - \mathrm{H}(K, C \mid T, X, X_K = C) \leq 3\delta_2. \tag{6}$$

**Definition 2.4.** Let $\nu$ denote the distribution of $T$ and let $\nu_0 := \nu \mid (X_K = C)$. For a specific transcript $t$, let $\rho_t$ denote the distribution $\mu \mid (T = t)$.

We can interpret (5) and (6) as bounding the expectations of appropriate functions of a random transcript distributed according to $\nu_0$. Inequality (2), though, is not of this form, since there is no conditioning on $(X_K = C)$; instead, it says

$$\mathbb{E}_{T \sim \nu} \left[ \Pr_{(X', K', C') \sim \rho_T} [\mathrm{out}(T) \neq \mathrm{AI}(X', K', C')] \right] \leq 3/\log^2 n. \tag{7}$$

Since we have $\Pr[X_K = C] = \frac{1}{2}$, every transcript $t$ satisfies $\nu_0(t) \leq 2\nu(t)$. Thus, switching the distribution in the outer expectation from $\nu$ to $\nu_0$ can at most double the left-hand side. In other words, we have

$$\mathbb{E}_{T_0 \sim \nu_0} \left[ \Pr_{(X', K', C') \sim \rho_{T_0}} [\mathrm{out}(T_0) \neq \mathrm{AI}(X', K', C')] \right] \leq 6/\log^2 n. \tag{8}$$

Finally, we can say that transcripts drawn from $\nu_0$ typically output "0", because

$$\Pr_{T_0 \sim \nu_0}[\operatorname{out}(T_0) \neq 0] \;=\; \Pr[\operatorname{out}(T) \neq \operatorname{AI}(X, K, C) \mid X_K = C] \;\leq\; 6/\log^2 n\,, \tag{9}$$

where the final step uses (2). By another averaging argument, applying Lemma 2.2 to the four inequalities (5), (6), (8) and (9), we conclude that there exists a transcript $t$ such that

$$\begin{aligned}
n - 1 - \operatorname{H}(X \mid K, C, X_K = C, T = t) &\leq 12\delta_1 n\,, \\
\log n - \operatorname{H}(K, C \mid X, X_K = C, T = t) &\leq 12\delta_2\,, \\
\Pr_{(X',K',C') \sim \rho_t}[\operatorname{out}(t) \neq \operatorname{AI}(X,K,C)] &\leq 24/\log^2 n\,, \text{ and} \\
\operatorname{out}(t) &= 0\,.
\end{aligned}$$

However, by the Fat Transcript Lemma (Lemma 2.6) below, it follows that no transcript can simultaneously satisfy the above four conditions. This completes the proof. □

At this point, we need to understand what is special about the distributions $\rho_t$ (from Definition 2.4), given that they arise from transcripts of private-coin communication protocols. The key fact we need here is the so-called *rectangle property* of deterministic communication protocols [14, Ch. 1]. More specifically, we need its extension to private-coin randomized protocols, as used, e.g., by Bar-Yossef et al. [4, Lemma 6.7].

However, there is a complication here due to the fact that Alice and Bob share some information. Had Bob *not* received any part of Alice's input, $\rho_t$ would have been a product of a distribution on values of $x$, and another distribution on values of $(k, c)$. But because Bob *does*, in fact, start out knowing $x_{1:\,k-1}$, we can only draw the weaker conclusion given in the following lemma.

**Lemma 2.5.** *Let $\mathcal{X} = \{0,1\}^n$ and $\mathcal{Y} = \{(w, k, c) \in \{0,1\}^* \times [n] \times \{0,1\} : |w| = k - 1\}$. Let $P$ be a private-coin protocol in which Alice receives a string $x \in \mathcal{X}$ while Bob receives $(w, k, c) \in \mathcal{Y}$, with the promise that $w = x_{1:\,k-1}$. Then, for every transcript $t$ of $P$, there exist functions $p_{A,t} : \mathcal{X} \to \mathbb{R}^+$ and $p_{B,t} : \mathcal{Y} \to \mathbb{R}^+$ such that*

$$\forall\, (x, k, c) \in \{0,1\}^n \times [n] \times \{0,1\} : \quad \rho_t(x, k, c) \;=\; p_{A,t}(x) \cdot p_{B,t}(x[1\,..\,k-1], k, c)\,.$$

*Proof.* Let $\mathcal{T}$ be the set of all possible transcripts of $P$ and let $T$ be a random transcript of $P$ on input $(X, K, C) \sim \mu$. By the rectangle property for private-coin protocols (Lemma 6.7 of [4]), there exist mappings $q_A : \mathcal{T} \times \mathcal{X} \to \mathbb{R}^+$ and $q_B : \mathcal{T} \times \mathcal{Y} \to \mathbb{R}^+$ such that

$$\Pr[T = t \mid (X, K, C) = (x, k, c)] \;=\; q_A(t; x) \cdot q_B(t; x_{1:k-1}, k, c)\,.$$

Recall that $\mu$ is just a uniform distribution. In particular, it decomposes as $\mu(x, k, c) = \mu_A(x)\mu_B(k, c)$. Thus, by Bayes' Theorem,

$$\begin{aligned}
\rho_t(x, k, c) &= \frac{\mu(x, k, c) \cdot \Pr[T = t \mid (X, K, C) = (x, k, c)]}{\Pr[T = t]} \\
&= \frac{\mu_A(x) \cdot \mu_B(k, c) \cdot q_A(t; x) \cdot q_B(t; x_{1:k-1}, k, c)}{\Pr[T = t]}\,.
\end{aligned}$$

Now set $p_{A,t}(x) := \mu_A(x) \cdot q_A(t; x)/\Pr[T = t]$ and $p_{B,t}(w, k, c) := \mu_B(k, c) \cdot q_B(t; w, k, c)$. □

We now state the promised lemma that, as shown above, finishes the proof of Theorem 2.3. We alert the reader that, *from here on, the distribution of $(X, K, C)$ is no longer uniform*; instead, we condition the uniform distribution on a specific transcript.

**Lemma 2.6** (Fat Transcript Lemma). *There exist positive real constants $\delta_3$ and $\delta_4$ such that, for every transcript $t$ of a private-coin communication protocol for AI, with $\operatorname{out}(t) = 0$, we have the following. Let $(X, K, C) \sim \rho_t$ and let $\varepsilon(n) = 24/\log^2 n$. Then the following conditions do not hold simultaneously:*

$$H(X \mid K, C, X_K = C) \geq (1 - \delta_3)n, \tag{10}$$
$$H(K, C \mid X, X_K = C) \geq \log n - \delta_4, \tag{11}$$
$$\mathbb{E}[\operatorname{AI}(X, K, C)] \leq \varepsilon(n). \tag{12}$$

*Proof.* Suppose, to the contrary, that (10), (11) and (12) do hold for every choice of $\delta_3$ and $\delta_4$. Since $C$ is determined by $X$ and $K$ whenever the condition $X_K = C$ holds, the left-hand side of (11) equals $H(K \mid X, X_K = C)$. Also, (12) is equivalent to $\Pr[X_K = C] \geq 1 - \varepsilon(n)$. Thus, we can simplify (11) to

$$H(K \mid X) \geq \Pr[X_K = C] \cdot H(K \mid X, X_K = C) \geq (1 - \varepsilon(n))(\log n - \delta_4) \geq \log n - 2\delta_4, \tag{13}$$

where the last step uses the bound $\varepsilon(n) = o(1/\log n)$. Similarly, we can simplify (10) to

$$H(X) \geq H(X \mid K, C) \geq \Pr[X_K = C] \cdot H(X \mid K, C, X_K = C) \geq (1 - \varepsilon(n))(1 - \delta_3)n \geq (1 - 2\delta_3)n. \tag{14}$$

We now expand (12). In what follows, we use notation of the form "$u0v$" to denote the concatenation of the string $u$, the length-1 string "0", and the string $v$.

$$\mathbb{E}[\operatorname{AI}(X, K, C)] = \sum_{k=1}^{n} \sum_{x \in \{0,1\}^n} \sum_{c \in \{0,1\}} \rho_t(x, k, c) \cdot \operatorname{AI}(x, k, c)$$
$$= \sum_{k=1}^{n} \sum_{u \in \{0,1\}^{k-1}} \sum_{b \in \{0,1\}} \sum_{v \in \{0,1\}^{n-k}} \sum_{c \in \{0,1\}} \rho_t(ubv, k, c) \cdot \operatorname{AI}(ubv, k, c). \tag{15}$$

Let $p_A = p_{A,t}$ and $p_B = p_{B,t}$ be the functions given by Lemma 2.5. Let $\lambda$ denote the distribution of $(X, K)$, i.e., let $\lambda(x, k) = \rho_t(x, k, 0) + \rho_t(x, k, 1)$. Now, noting that $\operatorname{AI}(ubv, k, c) = 1$ iff $b \neq c$, we can manipulate (15) as follows.

$$\mathbb{E}[\operatorname{AI}(X, K, C)] = \sum_{k=1}^{n} \sum_{u \in \{0,1\}^{k-1}} \sum_{v \in \{0,1\}^{n-k}} \Big(\rho_t(u0v, k, 1) + \rho_t(u1v, k, 0)\Big)$$
$$= \sum_{k=1}^{n} \sum_{u \in \{0,1\}^{k-1}} \sum_{v \in \{0,1\}^{n-k}} \Big(p_A(u0v) \cdot p_B(u, k, 1) + p_A(u1v) \cdot p_B(u, k, 0)\Big)$$
$$= \sum_{k=1}^{n} \sum_{u \in \{0,1\}^{k-1}} \Big(p_B(u, k, 1) \sum_{v \in \{0,1\}^{n-k}} p_A(u0v) + p_B(u, k, 0) \sum_{v \in \{0,1\}^{n-k}} p_A(u1v)\Big)$$
$$\geq \sum_{k=1}^{n} \sum_{u \in \{0,1\}^{k-1}} \Big(p_B(u, k, 0) + p_B(u, k, 1)\Big) \cdot \min\Big\{\sum_{v \in \{0,1\}^{n-k}} p_A(u0v), \sum_{v \in \{0,1\}^{n-k}} p_A(u1v)\Big\}$$
$$= \sum_{k=1}^{n} \sum_{u \in \{0,1\}^{k-1}} \min\Big\{\sum_{v \in \{0,1\}^{n-k}} \lambda(u0v, k), \sum_{v \in \{0,1\}^{n-k}} \lambda(u1v, k)\Big\}. \tag{16}$$

Let $\alpha : \{0,1\}^n \to [0,1]$ and $\beta : [n] \to [0,1]$ be the marginals of $\lambda$, i.e., $\alpha(x) := \sum_{k=1}^{n} \lambda(x, k)$ and $\beta(k) := \sum_{x \in \{0,1\}^n} \lambda(x, k)$. We now make the following crucial observations about these distributions.

**Claim 2.7.** *We have* $\|\lambda - \alpha \otimes \beta\|_1 = \sum_{x \in \{0,1\}^n} \sum_{k=1}^n |\lambda(x,k) - \alpha(x)\beta(k)| \leq \sqrt{(4 \ln 2) \cdot \delta_4}.$

*Proof.* Using the characterization of mutual information in terms of Kullback-Leibler divergence, we get

$$D_{KL}(\lambda \,\|\, \alpha \otimes \beta) \;=\; I(K:X) \;=\; H(K) - H(K \mid X) \;\leq\; 2\delta_4\,,$$

where the last step uses (13) and the basic fact that $H(K) \leq \log n$. The claim now follows from Pinsker's inequality (for which see, e.g., [9, Lemma 12.6.1]). □

**Claim 2.8.** *We have* $\sum_{k=1}^n |\beta(k) - 1/n| \leq \sqrt{(4 \ln 2) \cdot \delta_4}.$

*Proof.* Relax (13) to $H(K) \geq \log n - 2\delta_4$. Let $\gamma$ denote the uniform distribution on $[n]$. Then we have $D_{KL}(\beta \,\|\, \gamma) = \log n - H(K) \leq 2\delta_4$. Now apply Pinsker's inequality. □

Let $\delta_5 := \sqrt{(4 \ln 2) \cdot \delta_4}$. Using Claim 2.7 to estimate the expression (16), keeping in mind that any particular $\lambda(x,k)$ term appears at most once in the summation, we get

$$\mathbb{E}[\mathrm{AI}(X,K,C)] \;\geq\; \sum_{k=1}^n \beta(k) \sum_{u \in \{0,1\}^{k-1}} \min\left\{ \sum_{v \in \{0,1\}^{n-k}} \alpha(u0v), \sum_{v \in \{0,1\}^{n-k}} \alpha(u1v) \right\} - \delta_5. \quad (17)$$

For each $k \in [n]$, define the probability distribution $\hat{\alpha}_k$ on $\{0,1\}^{k-1}$ by $\hat{\alpha}_k(u) := \sum_{w \in \{0,1\}^{n-k+1}} \alpha(uw) = \Pr[X_{1:k-1} = u]$. Let $H_b : [0,1] \to [0,1]$ denote the binary entropy function, i.e., $H_b(z) := -z \log z - (1-z) \log(1-z)$. Let $H_b^{-1} : [0,1] \to [0, \frac{1}{2}]$ denote the (well-defined) inverse of this function. Observe that, if $Z$ is a binary random variable, then $\min\{\Pr[Z=0], \Pr[Z=1]\} = H_b^{-1}(H(Z))$. Using all this, we obtain

$$\min\left\{ \sum_{v \in \{0,1\}^{n-k}} \alpha(u0v), \sum_{v \in \{0,1\}^{n-k}} \alpha(u1v) \right\} \;=\; \hat{\alpha}_k(u) \cdot \min\left\{ \frac{\hat{\alpha}_{k+1}(u0)}{\hat{\alpha}_k(u)}, \frac{\hat{\alpha}_{k+1}(u1)}{\hat{\alpha}_k(u)} \right\}$$

$$=\; \hat{\alpha}_k(u) \cdot H_b^{-1}\Big( H(X_k \mid X_{1:k-1} = u) \Big). \quad (18)$$

Plugging this back into (17), we obtain

$$\mathbb{E}[\mathrm{AI}(X,K,C)] + \delta_5 \;\geq\; \sum_{k=1}^n \beta(k) \sum_{u \in \{0,1\}^{k-1}} \hat{\alpha}_k(u) \cdot H_b^{-1}\Big( H(X_k \mid X_{1:k-1} = u) \Big)$$

$$\geq\; \sum_{k=1}^n \beta(k) \cdot H_b^{-1}\left( \sum_{u \in \{0,1\}^{k-1}} \hat{\alpha}_k(u) \cdot H(X_k \mid X_{1:k-1} = u) \right) \quad (19)$$

$$=\; \sum_{k=1}^n \beta(k) \cdot H_b^{-1}\Big( H(X_k \mid X_{1:k-1}) \Big) \;\geq\; H_b^{-1}\left( \sum_{k=1}^n \beta(k) \cdot H(X_k \mid X_{1:k-1}) \right) \quad (20)$$

$$\geq\; H_b^{-1}\left( \sum_{k=1}^n \frac{1}{n} \cdot H(X_k \mid X_{1:k-1}) - \delta_5 \right) \;=\; H_b^{-1}\left( \frac{H(X)}{n} - \delta_5 \right), \quad (21)$$

where (19) and (20) follow from Jensen's inequality (and the convexity of $H_b^{-1}$) and (21) uses Claim 2.8 and the fact that $H_b^{-1}$ is increasing on $[0,1]$. We now invoke (14) and (12) to obtain

$$\varepsilon(n) + \delta_5 \;\geq\; H_b^{-1}(1 - 2\delta_3 - \delta_5)\,.$$

Recall that $\delta_5 = \sqrt{(4 \ln 2) \cdot \delta_4}$. By choosing $\delta_3$ and $\delta_4$ small enough, we can make the left-hand side of the above inequality approach 0 and the right-hand side approach $\frac{1}{2}$, and we finally have our contradiction. □

# 3  A Direct Sum Argument

Let MULTI-AI$_{m,n}$ denote the following communication problem, involving $2m$ players $A_1, B_1, \ldots, A_m, B_m$. Each $A_i$ receives a string $x^i \in \{0,1\}^n$ and each $B_i$ receives an integer $k^i \in [n]$, a bit $c^i \in \{0,1\}$, and the length-$(k^i - 1)$ prefix $x^i_{1:k^i-1}$ of $x^i$. The players wish to compute the predicate $\bigvee_{i=1}^m \text{AI}_n(x^i, k^i, c^i)$. The players may use private random strings and a common public random string, and use $p$ *rounds*, where each round consists of a player sending an $s$-bit message privately to the next player in the following sequence:

$$A_1 \to B_1 \to A_2 \to B_2 \to \cdots\cdots \to A_m \to B_m \to A_m \to A_{m-1} \cdots\cdots \to A_1\,.$$

At the end of these $p$ rounds, $A_1$ must announce the answer, which is required to be correct with probability at least $(1-\varepsilon)$ on each possible input. Call such a protocol a $[p, s, \varepsilon]$-protocol. We then have:

**Theorem 3.1.** *Every $[p, s, 1/\log^2 n]$-protocol for MULTI-AI$_{m,n}$ satisfies $ps = \Omega\left(\min\{m, n\}\right)$.*

This theorem is easily seen to be near-optimal: even with $p = 1$, we have a trivial protocol achieving $s = O(n)$ and another trivial protocol achieving $s = O(m \log n)$.

Notice that the augmented index problem studied in Section 2 satisfies AI$_n$ = MULTI-AI$_{1,n}$. Intuitively, a protocol for MULTI-AI$_{m,n}$ must solve $m$ independent AI instances, and thus, must use about $m$ times the communication that a single instance requires. To prove Theorem 3.1, we formalize this intuition as a direct sum theorem, which we can prove using a suitable refinement of the information complexity paradigm [6]. To state this direct sum theorem, we need a suitable notion of information cost for protocols solving MULTI-AI. Let $Q$ be a $[p, s, \varepsilon]$-protocol for MULTI-AI$_{m,n}$. Let $\xi$ be a distribution on inputs to $Q$ and let $M_m$ denote the sequence of messages sent by player $B_m$ when $Q$ is run on a random input $\langle (X^i, K^i, C^i) \rangle_{i=1}^m \sim \xi$, using a public random string $R$. We strategically define the *information cost* of $Q$ under $\xi$ to be

$$\text{icost}_\xi(Q) := \text{I}(M_m : K^1, C^1, \ldots, K^m, C^m \mid X^1, \ldots, X^m, R)\,. \tag{22}$$

It is worth noting that when $m = 1$, i.e., we are considering a protocol for AI, this definition specializes to that of $\text{icost}^B_\xi(Q)$ in (1). This is proved in Lemma A.1.

**Theorem 3.2** (Direct sum theorem for AI). *Suppose there exists a $[p, s, \varepsilon]$-protocol $Q$ for MULTI-AI$_{m,n}$. Then there exists an $\varepsilon$-error randomized protocol $P$ for AI in which Alice sends at most $ps$ bits in total, such that $m \cdot \text{icost}^B_{\mu_0}(P) \le \text{icost}_{\mu_0^{\otimes m}}(Q)$, where $\mu_0$ is as in Definition 2.1 and $\mu_0^{\otimes m}$ denotes the $m$-fold product of $\mu_0$ with itself.*

*Proof.* This is a straightforward generalization, to multiple rounds, of a similar theorem of Magniez et al. [15], which applied only to restricted families of one-round protocols. Details appear in Appendix A. □

We can now prove our multi-round communication lower bound on MULTI-AI as follows.

*Proof of Theorem 3.1.* Let $Q$ be a $[p, s, 1/\log^2 n]$-protocol for MULTI-AI$_{m,n}$. From basic information theory, it follows that $\text{icost}_{\mu_0^{\otimes m}}(Q) \le ps$. Now, by Theorem 3.2, there exists a protocol $P$ for AI with $\text{icost}^B_{\mu_0}(P) \le ps/m$ and in which Alice communicates at most $ps$ bits, so that $\text{icost}^A_{\mu_0}(P) \le ps$. By Theorem 2.3, either $ps/m = \Omega(1)$ or $ps = \Omega(n)$; i.e., $ps = \Omega\left(\min\{m, n\}\right)$. □

# 4  Streaming Language Recognition and Passive Memory Checking

In this section we present our results for recognizing certain languages in the data stream model. Of particular interest is DYCK(2), the language consisting of the strings of well-balanced parentheses in two types of parentheses. Formally, representing '(', ')', '[', and ']' as $a$, $\bar{a}$, $b$, and $\bar{b}$ respectively,

**Definition 4.1.** DYCK(2) is the language generated by the context-free grammar $S \to aS\bar{a} \mid bS\bar{b} \mid SS \mid \epsilon$.

An important class of memory checking problems, which we call *passive checking*, can also be viewed as language recognition problems in the data stream model. For example, we define PQ to be the language corresponding to transcripts of operations, or "interaction sequences," of a priority queue that begins and ends with an empty queue. (Without this restriction, the resulting language would require $\Omega(N)$ space to recognize, for simple reasons [8, Theorem 4].) Formally,

**Definition 4.2.** An *interaction sequence* $\sigma = \sigma_1\sigma_2\ldots\sigma_N$ is a string over the alphabet $\Sigma = \{\text{ins}(u), \text{ext}(u) : u \in [U]\}$. Let PQ = PQ($U$) be the language defined over $\Sigma$ where $\text{ins}(u)$ is interpreted as an insertion of $u$ into a priority queue, and $\text{ext}(u)$ as an extraction of $u$ from the priority queue. The state of the queue at any step $j$ can be represented by a multiset $M_j$ so that

$$M_0 = \emptyset; \quad M_j = M_{j-1} \setminus \{\min(M_{j-1})\} \text{ if } \sigma_j = \text{ext}(v); \text{ and } \quad M_j = M_{j-1} \cup \{u\} \text{ if } \sigma_j = \text{ins}(u). \tag{23}$$

Then $\sigma \in$ PQ for $|\sigma| = N$ iff $M_N = \emptyset$ and $\forall j \in [N]\ (\sigma_j = \text{ext}(u) \Rightarrow u = \min(M_{j-1}))$.

We start by showing that a recognizer for PQ can also recognize DYCK(2) via an online transformation. The reduction proceeds as follows. Consider a string $p$ over parentheses $\{a, \bar{a}, b, \bar{b}\}$ and define

$$\text{height}(p) := |\{j : p_j \in \{a, b\}\}| - |\{j : p_j \in \{\bar{a}, \bar{b}\}\}|$$

and $\text{height}(\epsilon) = 0$. We transform $p$ into $\psi(p) = \phi(p_{1:1})\phi(p_{1:2})\ldots\phi(p_{1:N})$ where:

$$\phi(p_{1:i}) = \begin{cases} \text{ins}(2N - 2\,\text{height}(p_{1:i-1})) & \text{if } p_i = a \\ \text{ext}(2N - 2\,\text{height}(p_{1:i})) & \text{if } p_i = \bar{a} \\ \text{ins}(2N - 2\,\text{height}(p_{1:i-1}) - 1) & \text{if } p_i = b \\ \text{ext}(2N - 2\,\text{height}(p_{1:i}) - 1) & \text{if } p_i = \bar{b} \end{cases}$$

For example, the string $\langle a, a, \bar{a}, b, \bar{b}, \bar{a}\rangle$ is transformed into $\langle \text{ins}(12), \text{ins}(10), \text{ext}(10), \text{ins}(9), \text{ext}(9), \text{ext}(12)\rangle$. The proof that $\psi(p) \in$ PQ if and only if $p \in$ DYCK(2) is given in Appendix B.

**Lemma 4.3.** *There exists an $O(\log N)$-space stream reduction from DYCK(2) to PQ(4N).*

Our first result on the complexity of stream language recognition uses Theorem 3.1 to resolve the conjecture of Magniez, et al. [15] regarding the multi-pass complexity of DYCK(2) and PQ.

**Theorem 4.4** (Multi-pass Lower Bounds for DYCK and PQ). *Let $L$ denote either DYCK(2) or PQ($N$). Suppose there exists a $O(1/\log^2 N)$-error, p-pass, s-space, randomized streaming algorithm that recognizes $L$ on length-$N$ streams. Then $ps = \Omega(\sqrt{N})$.*

*Proof.* Using the reduction of Magniez, et al. [15], an $\varepsilon$-error $p$-pass randomized streaming algorithm for DYCK(2) that uses $s$ bits of space on streams of length $\Theta(mn)$ can be turned into a $[p, s, \varepsilon]$-protocol for MULTI-AI$_{m,n}$. One can similarly reduce MULTI-AI$_{m,n}$ to PQ($N$); this was implicitly claimed without proof in [15]. Alternatively, Lemma 4.3 gives an explicit reduction from MULTI-AI$_{m,n}$ to PQ via DYCK(2). To complete the proof, we combine these reductions with Theorem 3.1, setting $m = n$. □

**Unidirectional versus Bidirectional Passes.** As noted earlier, DYCK(2) can be recognized in $O(\log^2 N)$ space using two passes, one in each direction. On the other hand, the above theorem implies that achieving polylog($n$) space with only unidirectional access to the input would require $\widetilde{\Omega}(\sqrt{N})$ passes. To the best of our knowledge, this is the first explicit demonstration of such a strong separation between these two natural data stream models.

**Algorithm 1** PQ-CHECK

1: **input** $\sigma = \sigma^{E_1}\sigma^{I_1}\sigma^{E_2}\sigma^{I_2}\ldots\sigma^{E_r}\sigma^{I_r}$ where $\sigma^{E_1} = \sigma^{I_r} = \emptyset$
2: **for** $k \in \{1,\ldots,r\}, u \in \{1,\ldots,U\}$, **do** $f[k] \leftarrow 0, X[k,u] \leftarrow 0, Y[k,u] \leftarrow 0, Z[k,u] \leftarrow 0$
3: **for** $i \in \{1,\ldots,r\}$ **do**
4:     **for** $\text{ext}(u) \in \sigma^{E_i}$ **do**
5:         $\ell \leftarrow \min\{k : f[k] \leq u\}$
6:         $Y[\ell, u] \leftarrow Y[\ell, u] + 1$
7:         $Z[\ell, u] \leftarrow \max(Y[\ell, u], Z[\ell, u])$
8:         **for** $1 \leq k < i$ **do** $f[k] \leftarrow \max(u, f[k])$
9:     **end for**
10:     **for** $\text{ins}(u) \in \sigma^{I_i}$ **do**
11:         $\ell \leftarrow \min\{k : f[k] \leq u\}$
12:         **if** $f[\ell] < u$ **then** $X[\ell, u] \leftarrow X[\ell, u] + 1$
13:         **if** $f[\ell] = u$ **then** $Y[\ell, u] \leftarrow Y[\ell, u] - 1$
14:     **end for**
15: **end for**
16: **if** $X \neq Z$ or $X \neq Y$ **then** reject **else** accept

## 4.1 Passive Checking of Priority Queues

Given the connection between PQ and DYCK(2) shown in Lemma 4.3, one might hope to adapt the algorithms of [15] to this problem. However, there seems to be no such easy reduction in this direction. For intuition, observe that DYCK(2) has a much stricter requirement on the permitted strings: if its second half consists of close-parentheses only, then its first half is uniquely determined. On the other hand, in PQ, one can find $(N/2)!$ sequences consisting of $N/2$ insertions followed by $N/2$ extractions that all agree on the second half. This suggests that the two languages are quite different.

We therefore give a novel algorithm that leads to the following theorem, which is the main upper bound result in this paper.

**Theorem 4.5.** *We can recognize the language PQ in one pass, using $O(\sqrt{N}(\log U + \log N))$ bits of space: an input $\sigma \in$ PQ is accepted with certainty, and an input $\sigma \notin$ PQ is rejected with probability $\geq 1 - 1/N^2$.*

**Overview of the Algorithm.** We first present a $O(Ur(\log U + \log N))$ space algorithm for the case when the input string can be decomposed as $\sigma = \sigma^{E_1}\sigma^{I_1}\sigma^{E_2}\sigma^{I_2}\ldots\sigma^{E_r}\sigma^{I_r}$ where $\sigma^{E_i}$ is a sequence of extracts and $\sigma^{I_i}$ is a sequence of inserts. We refer to $\sigma^{E_i}\sigma^{I_i}$ as the $i$th *epoch* of the string and note that, for sufficiently large $r$, any $\sigma$ is of this form. After presenting the full space algorithm, we show how to transform $\sigma$ such that $r = O(\sqrt{N})$ and subsequently, to reduce the space to $\tilde{O}(\sqrt{N})$. Finally, a necessary condition for $\sigma \in$ PQ is that the extracts in each $\sigma^{E_i}$ are in ascending order and that $\sigma^{E_1} = \sigma^{I_r} = \emptyset$. Since both conditions are easily verified, we assume that they are satisfied.

We present the algorithm PQ-CHECK as Algorithm 1. We first describe its properties informally, before proceeding to a more formal analysis.

1. For each epoch $k$, PQ-CHECK maintains a value $f[k]$ that is the maximum value that has been extracted after the $k$th epoch. In particular, at the very start of the $i$th epoch, $f[i-1] = 0$.

2. Each insert/extract of $u$ is *assigned* to the earliest epoch "consistent" with the current $f$ values maintained by PQ-CHECK, i.e., $\ell = \min\{k : f[k] \leq u\}$. Each $\text{ext}(u) \in \sigma^{E_i}$ is assigned to an epoch between 1 and $i - 1$ (this follows because the extracts in $\sigma^{E_i}$ are in increasing order and $f[i-1]$ equals 0 when the first extract in $\sigma^{E_i}$ is processed), while each $\mathrm{ins}(u) \in \sigma^{I_i}$ is assigned to an epoch between 1 and $i$. Importantly, for $\sigma \in \mathrm{PQ}$, each $\mathrm{ext}(u)$ will be assigned to the same epoch as the most recent $\mathrm{ins}(u)$.

3. The algorithm maintains arrays $X, Y$, and $Z$ to track information about occurrences of item $u$ assigned to epoch $k$ (we later use hashing techniques to reduce the size of this information). Informally, $X$ tracks the number of insertions of $u$ assigned to epoch $k$ before the first extraction of $u$ that is assigned to epoch $k$, while $Y$ tracks the number of extractions of $u$ assigned to epoch $k$ minus the number of insertions of $u$ assigned to epoch $k$ from the first extraction of $u$ assigned to epoch $k$ onwards. A necessary condition is that these two counts should agree. However, this counting alone fails to detect extractions of $u$ that appear before the corresponding insertions. Therefore, $Z$ is used to identify the maximum "balance" of $u$ during epoch $k$. This should also match $X$ if the sequence is correct, and we later show that these are sufficient conditions to check membership in PQ.

Define $f_t(k) = \max\{u : \sigma_i = \mathrm{ext}(u), |\sigma^{E_1} \ldots \sigma^{I_k}| + 1 \leq i \leq t\}$. For $u \in [U]$ and $t \in [N]$, define $b(t, u) = \min\{k : f_t(k) \leq u\}$. Given an interaction sequence $\sigma$ and $u \in [U]$, define

$$\mathrm{cnt}(\sigma, u) := |\{t : \sigma_t = \mathrm{ins}(u)\}| - |\{t : \sigma_t = \mathrm{ext}(u)\}|.$$

**Lemma 4.6.** *After processing the $t$th element, Algorithm 1 has computed $f[k] = f_t(k)$, i.e., the maximum value extracted after the end of the $k$th epoch. For all $k$, $f[k]$ is non-decreasing as $t$ increases.*

*Proof.* Observe that Algorithm 1 only updates $f[k]$ in Line 8, for $k < i$ where the current epoch is the $i$th epoch. The equivalence of $f[k]$ and $f_t(k)$ follows immediately by an inductive argument over $t$. $f[k] = f_t(k)$ is seen to be non-decreasing by inspection of the definition of $f_t(k)$. $\square$

**Lemma 4.7.** *Let $X_t(k, u), Y_t(k, u)$ and $Z_t(k, u)$ denote the values of $X[k, u], Y[k, u]$, and $Z[k, u]$ after processing the $t$th element. Assume that the first $t$ elements of the interaction sequence are a prefix of some interaction sequence in PQ, i.e., for all $j \in [t], (\sigma_j = \mathrm{ext}(v) \implies v = \min(M_{j-1}))$ where $\{M_j\}_{j=0}^N$ is the family of multisets defined in Eq. (23). Then, for any $u \in [U]$ and $k = b(t, u)$, we have:*

$$\mathrm{cnt}(\sigma_{1:t}, u) = X_t(k, u) - Y_t(k, u)$$

*and for $k < b(t, u)$, $X_t(k, u) = Y_t(k, u)$.*

*Proof.* Let $u \in [U]$ be an arbitrary element. We proceed by induction on $t$. The lemma is true for $t = 0$ where $X_0(k, u) = Y_0(k, u) = 0$ for all $k, u$. For the inductive step with $k = b(t-1, u)$, there are four cases to consider:

1. Case $\sigma_t = \mathrm{ins}(u)$: In this case $b(t-1, u) = b(t, u) = k$. Therefore,

   $$\mathrm{cnt}(\sigma_{1:t}, u) = \mathrm{cnt}(\sigma_{1:t-1}, u) + 1 = 1 + X_{t-1}(k, u) - Y_{t-1}(k, u) = X_t(k, u) - Y_t(k, u)$$

   The last step follows whether or not $f_t(b(t-1, u)) = u$ (lines 12 and 13 in Algorithm 1).

2. Case $\sigma_t = \mathrm{ext}(u)$: In this case $b(t-1, u) = b(t, u) = k$. Therefore,

   $$\mathrm{cnt}(\sigma_{1:t}, u) = \mathrm{cnt}(\sigma_{1:t-1}, u) - 1 = X_{t-1}(k, u) - (Y_{t-1}(k, u) + 1) = X_t(k, u) - Y_t(k, u)$$

3. Case $\sigma_t = \mathrm{ins}(v)$ for $v \neq u$ or $\sigma_t = \mathrm{ext}(v)$ for $v < u$: In this case $b(t-1, u) = b(t, u) = k$. Therefore,

   $$\mathrm{cnt}(\sigma_{1:t}, u) = \mathrm{cnt}(\sigma_{1:t-1}, u) = X_{t-1}(k, u) - Y_{t-1}(k, u) = X_t(k, u) - Y_t(k, u)$$

4. Case $\sigma_t = \text{ext}(v)$ for $u < v$. In this case we know that $\text{cnt}(\sigma_{1:t-1}, u) = 0$. Assume it was not: then either there is a minimal prefix of $\sigma$ for some $j$ such that $\text{cnt}(\sigma_{1:j}, u) < 0$ which implies that $\sigma_j = \text{ext}(u)$ but $u \neq \min(M_{j-1})$; or else $\text{cnt}(\sigma_{1:t-1}, u) > 0$ which implies that $v \neq \min(M_{t-1})$ since $\min(M_{t-1}) \leq u < v$. Either way, we contradict our assumption on $\sigma$. Therefore,

$$\text{cnt}(\sigma_{1:t}, u) = \text{cnt}(\sigma_{1:t-1}, u) = X_{t-1}(b(t-1, u), u) - Y_{t-1}(b(t-1, u), u)$$
$$= X_t(b(t-1, u), u) - Y_t(b(t-1, u), u)$$

If $b(t-1, u) = b(t, u)$ we are done but it is possible that $b(t-1, u) \neq b(t, u)$. This is because following this extraction, for all $1 \leq \ell < i$, we set $f[\ell]$ to $\max(f[\ell], v) > u$ which forces $b(t, u) = i$, where $i$ is the current epoch. But then $X_t(b(t, u), u) = Y_t(b(t, u), u) = 0$ since no inserts or extracts of $u$ can yet have been associated with epoch $i$. Hence, even if $b(t-1, u) \neq b(t, u)$, $\text{cnt}(\sigma_{1:t}, u) = X_t(k, u) - Y_t(k, u)$ for $k = b(t, u)$.

In all cases, for $k < b(t-1, u)$, we observe that Algorithm 1 does not modify $X[k, u]$ or $Y[k, u]$ and these are already equal by the induction hypothesis. If $k = b(t-1, u) < b(t, u)$, then, as reasoned in case 4 above, we have $X_t(k, u) = Y_t(k, u)$ as required. □

**Theorem 4.8.** *If $\sigma \notin$ PQ, Algorithm 1 rejects, else it accepts.*

*Proof.* If $\sigma \notin$ PQ, consider the minimum $t$ such that $\sigma_t = \text{ext}(u)$ and $u \neq \min(M_{t-1})$. Let $k = b(t-1, u)$. There are two possibilities. First, suppose $u \notin M_{t-1}$. Then, by Lemma 4.7, before processing $\sigma_t$, $X_{t-1}(k, u) - Y_{t-1}(k, u) = 0$. After processing $\sigma_t$ we have $Y_t(k, u) = Y_{t-1}(k, u) + 1$. Hence,

$$Z_t(k, u) \geq Y_t(k, u) > X_t(k, u).$$

Since $Z_s(k, u)$ is non-decreasing in $s$ and $X_s(k, u) = X_t(k, u)$ for $s > t$ after $f(k)$ becomes equal to $u$, at the end of the algorithm $Z_N(k, u) \neq X_N(k, u)$. Hence the algorithm rejects $\sigma$. Otherwise, suppose $u \in M_{t-1}$ but $\min(M_{t-1}) = v \neq u$. Then $\text{cnt}(\sigma_{1:t-1}, v) > 0$. Let $k = b(t-1, v)$ and by Lemma 4.7, $X_{t-1}(k, v) - Y_{t-1}(k, v) > 0$. Once $\text{ext}(u)$ is processed, $f[k]$ is increased to $u$ and hence $X_s(k, v) > Y_s(k, v)$ for all $s > t$, and the algorithm rejects.

If $\sigma \in$ PQ, then by Lemma 4.7, at $t = N$, $X_t(k, u) - Y_t(k, u) = 0$ for all $u, k$. Consequently, $Z_t(k, u) \geq Y_t(k, u) = X_t(k, u)$ for all $k, u$. Since $\text{cnt}(\sigma_t, u) \geq 0$ for any $\sigma \in$ PQ, $Y_t(k, u) \leq X_t(k, u)$ for all $t$. Hence $Z_t(k, u) \leq X_t(k, u)$ and so $X_N = Y_N = Z_N$ and the algorithm accepts. □

**Local Consistency.** We now consider a substring $\sigma'$ of $\sigma$ and show that if it does not violate some local conditions, then without loss of generality it can be assumed to be in a specific form.

**Definition 4.9.** We say $\sigma'$ is *locally consistent* if both

1. $\forall i < k, u < v : (\sigma'_i = \text{ins}(u)) \wedge (\sigma'_k = \text{ext}(v)) \implies (\text{cnt}(\sigma'_{i+1:k-1}, u) < 0)$.

2. $\forall i < k, u > v : (\sigma'_i = \text{ext}(u)) \wedge (\sigma'_k = \text{ext}(v)) \implies (\text{cnt}(\sigma'_{i+1:k-1}, v) > 0)$.

Observe that if $\sigma'$ is not locally consistent, then $\sigma \notin$ PQ, since the identified subsequence includes an extraction of an item which cannot be the smallest in the priority queue.

**Lemma 4.10.** *Given $\sigma = \sigma^{\text{pref}}\sigma'\sigma^{\text{suff}}$. If $\sigma'$ is locally consistent, then there exists a mapping $\gamma(\sigma') = \sigma^a\sigma^b\sigma^c\sigma^d$ such that $\sigma^{\text{pref}}\sigma'\sigma^{\text{suff}} \in$ PQ iff $\sigma^{\text{pref}}\gamma(\sigma')\sigma^{\text{suff}} \in$ PQ. Here, $\sigma^a$ and $\sigma^c$ are both sequences of extracts in increasing order; and $\sigma^b$ and $\sigma^d$ are both sequences of inserts. The algorithm* SUB-CHECK *tests if $\sigma'$ is locally consistent and, if so, computes $\gamma(\sigma')$ in time $O(|\sigma'| \log |\sigma'|)$.*

**Algorithm 2** SUB-CHECK

1: **input** $\sigma'$
2: $f \leftarrow 0; w \leftarrow 0; E \leftarrow \emptyset; I \leftarrow \{\infty\}$
3: **for** $i \in [|\sigma'|]$ **do**
4:    **if** $\sigma'_i = \text{ins}(u)$ **then** $I \leftarrow I \cup \{u\}$
5:    **if** $\sigma'_i = \text{ext}(v)$ **then**
6:       $m \leftarrow \min(I)$
7:       **if** $(v > m)$ **then** reject
8:       **if** $(v = m)$ **then** $I \leftarrow I \setminus \{v\}; w \leftarrow \max(w, v)$
9:       **if** $(v < m)$ **then**
10:          **if** $v < \max(f, w)$ **then** reject
11:          $f \leftarrow v; E \leftarrow E \cup \{v\}$
12:       **end if**
13:    **end if**
14: **end for**
15: **output** $\langle \text{ext}(v_1), \ldots, \text{ext}(v_{|E|}), \text{ins}(w), \text{ext}(w), \text{ins}(u_1), \ldots, \text{ins}(u_{|I|}) \rangle$ where $v_i$ and $u_i$ are the $i$th smallest values of $E$ and $I$ respectively

*Proof.* We first define the mapping $\gamma$ procedurally based on local rearrangements of the locally consistent $\sigma'$ which maintain local consistency. First consider all adjacent character pairs of the form $\text{ins}(u), \text{ext}(v)$. Since the string is locally consistent, $u \geq v$. Whenever $u > v$, we interchange these characters to obtain $\text{ext}(v), \text{ins}(u)$, without affecting local consistency or membership in PQ. Hence, we may assume that for every two adjacent characters $\text{ins}(u), \text{ext}(v)$, we have $u = v$, i.e., the pair represents an insertion followed immediately by an extraction of the same item. This generates a string $\alpha(\sigma')$ that satisfies $\sigma^{\text{pref}}\alpha(\sigma')\sigma^{\text{suff}} \in$ PQ iff $\sigma^{\text{pref}}\sigma'\sigma^{\text{suff}} \in$ PQ.

We next define two rearrangement rules on substrings of length three in $\alpha(\sigma')$. If applied to a string that was not locally consistent, they could "fix" errors, and lead to strings which are in PQ; however, since they are applied to locally consistent strings, the rearrangement preserves membership in PQ.

1. $\text{ins}(u) \text{ext}(u) \text{ext}(v) \to \text{ext}(v) \text{ins}(u) \text{ext}(u)$.

2. $\text{ins}(v) \text{ins}(u) \text{ext}(u) \to \text{ins}(u) \text{ext}(u) \text{ins}(v)$.

By repeatedly applying these rearrangement rules to $\alpha(\sigma')$ until no further rearrangement is possible we obtain $\beta(\sigma')$. Define the potential function $\Phi$ over interaction sequences as $\Phi(\sigma) = \sum_{\sigma_i = \text{ext}(u)} i$. Observe that each rearrangement reduces $\Phi$ by 1, so the process terminates. Let $\beta(\sigma')$ denote the final permutation and note that $\sigma^{\text{pref}}\beta(\sigma')\sigma^{\text{suff}} \in$ PQ iff $\sigma^{\text{pref}}\sigma'\sigma^{\text{suff}} \in$ PQ. Then, for some $t_1, t_2, t_3$, $\beta(\sigma')$ has the form,

$$\langle \text{ext}(v_1), \ldots, \text{ext}(v_{t_1}), \text{ins}(w_1), \text{ext}(w_1), \text{ins}(w_2), \text{ext}(w_2), \ldots, \text{ins}(w_{t_2}), \text{ext}(w_{t_2}), \text{ins}(u_1), \ldots, \text{ins}(u_{t_3}) \rangle$$

where $v_1 \leq v_2 \leq \ldots \leq v_{t_1}$. For $w = \max\{w_1, \ldots, w_{t_2}\}$, define $\gamma(\sigma') = \sigma^a \sigma^b \sigma^c \sigma^d$ where

$$\sigma^a = \langle \text{ext}(v_1), \ldots, \text{ext}(v_{t_1}) \rangle, \quad \sigma^b = \langle \text{ins}(w) \rangle, \quad \sigma^c = \langle \text{ext}(w) \rangle, \quad \text{and} \quad \sigma^d = \langle \text{ins}(u_1), \ldots, \text{ins}(u_{t_3}) \rangle.$$

Observe that $\sigma^{\text{pref}}\gamma(\sigma')\sigma^{\text{suff}} \in$ PQ iff $\sigma^{\text{pref}}\beta(\sigma')\sigma^{\text{suff}} \in$ PQ and $\sigma^{\text{pref}}\sigma'\sigma^{\text{suff}} \in$ PQ iff $\sigma^{\text{pref}}\gamma(\sigma')\sigma^{\text{suff}} \in$ PQ.

We next show that it is possible to test local consistency and compute $\gamma(\sigma')$ directly in $O(|\sigma'| \log |\sigma'|)$ time. Consider SUB-CHECK in Algorithm 2.

We first argue that Algorithm 2 correctly determines whether $\sigma'$ is locally consistent. First observe that $I$ records the multiset of items which have been inserted in $\sigma'$ and not yet extracted. A violation of Condition 1 in Definition 4.9 is detected in line 7 where the existence of $m \in I$ with $m < v$ indicates that an insufficient number of $\text{ext}(m)$ have occurred before the $\text{ext}(v)$ being considered.

A violation of Condition 2 is detected in line 10: this is when the current character is $\text{ext}(v)$ but there was an $\text{ext}(u)$ for $u > v$ earlier but no subsequent $\text{ins}(v)$ that could be matched with the current $\text{ext}(v)$. This is monitored via two variables, $f$ and $w$. $w$ is the maximum value extracted that is matched to an insertion happening within $\sigma'$. $f$ is the most recent value extracted that is not matched within $\sigma'$. Observe that because of the test in line 10, $f$ is non-decreasing. Consequently, $\max(f, w)$ is the largest value extracted so far. If there is some $u > v$ such that $\text{ext}(u)$ occurs in $\sigma'$ before $\text{ext}(v)$, then $\max(f, w) \geq u > v$. Hence, it suffices to track only the greatest extracted item in $\sigma'$. We can be sure that there is no $\text{ins}(v)$ matching the $\text{ext}(v)$ since $m = \min(M_i) > v$: if $v$ were matched, it would be present in $I$ and found as $m$.

The algorithm computes $\gamma$ correctly: $I$ is the multiset of items that are inserted but not extracted in $\sigma$, and $E$ is the multiset of items that are extracted without a matching insert in $\sigma'$. As noted above, $w$ tracks the greatest item which is inserted and subsequently extracted in $\sigma'$, so the output has the necessary form. Implementing $I$ and $E$ as priority queues means that each character is processed in $O(\log |\sigma'|)$ time, giving total $O(|\sigma'| \log |\sigma'|)$ time and $O(|\sigma'|)$ space. □

Consequently, by breaking $\sigma$ into sequential substrings of length $l$ and reordering each substring (unless we determine the substring is not locally consistent) we may ensure that the interaction sequence has the form $\sigma = \sigma^{E_1} \sigma^{I_1} \sigma^{E_2} \sigma^{I_2} \ldots \sigma^{E_r} \sigma^{I_r}$ where $r = 2\lceil N/l \rceil$. The final algorithm runs PQ-CHECK and SUB-CHECK in parallel. The space required by SUB-CHECK is $O(l \log U)$ bits and we will show that PQ-CHECK can be implemented in $O(r(\log N + \log U))$ bits. Setting $l = \sqrt{N}$ yields Theorem 4.5.

**Finishing the Proof: A Small-Space Implementation of PQ-CHECK.** Rather than maintain the arrays $X, Y$, and $Z$ explicitly in PQ-CHECK, it suffices to keep a linear hash (which serves as a homomorphic fingerprint) of each array. These fingerprints can be compared, and if they match in Line 16, then, with high probability, the arrays agree. In Line 7 we need to perform a max operation between two values. This can be done by maintaining $Y[k, f_t(k)]$ and $Z[k, f_t(k)]$ explicitly for each $k$. At any time, there are at most $r$ such values that are needed: observe that when $f_{t+1}(k) > f_t(k)$, $Y[k, f_t(k)]$ and $Z[k, f_t(k)]$ are never subsequently altered. The new values for $Y[k, f_{t+1}(k)]$ and $Z[k, f_{t+1}(k)]$ are initialized to 0. Hence, the space of the algorithm is $O(r)$ words to store the $Y[k, f[k]]$, $Z[k, f[k]]$ and $f[k]$ values, and a constant number of fingerprints to represent $X, Y$, and $Z$.

## 4.2 Passive Checking of Stacks, Queues, and Deques

**Stack.** Let STACK denote the language over interaction sequences that corresponds to stack operations. Now $\text{ins}(u)$ corresponds to an insertion of $u$ to a stack, and $\text{ext}(u)$ is an extraction of $u$ from the stack. Then $\sigma \in$ STACK iff $\sigma$ corresponds to a valid transcript of operations on a stack which starts and ends empty. That is, the state of the stack at any step $j$ can be represented by a string $S^j$ so that $S^0 = \emptyset$, $S^j = uS^{j-1}$ if $\sigma_j = \text{ins}(u)$ and $S^j = S^{j-1}_{2:|S^{j-1}|}$ if $\sigma_j = \text{ext}(u)$. Then $\sigma \in$ STACK for $|\sigma| = N$ iff

$$S^N = \emptyset \quad \text{and} \quad \forall j \in [N], (\sigma_j = \text{ext}(u) \implies u = S^{j-1}_1)$$

**Theorem 4.11.** *Every $O(1/\log^2 N)$-error, $p$-pass, $s$-space randomized streaming algorithm to recognize* STACK *on length $N$ streams must satisfy $ps = \Omega(\sqrt{N})$. It is possible to recognize* STACK *in one pass with $O(\sqrt{N} \log N)$ bits of space with high probability.*

*Proof.* First, we observe that for $U = 2$, DYCK$(U)$ = STACK if we associate $\text{ins}(u)$ with $u$ and $\text{ext}(u)$ with $\bar{u}$. Therefore, the lower bound follows immediately. For the upper bound, the one-pass algorithm from [15] to recognize DYCK(2) can be used to recognize STACK over arbitrary $U$ by appealing to their reduction from DYCK$(U)$ to DYCK(2). □

We note that the algorithm of [15] for recognizing DYCK(2) can be used directly to recognize DYCK($U$) rather than appealing to the reduction from DYCK($U$) to DYCK(2). In outline, the algorithm works as follows. The input string is broken into blocks of length $\sqrt{N}$. Within each block, any adjacent pair of the form $\langle\text{ins}(u), \text{ext}(u)\rangle$ can be matched and removed. When no further removals of pairs are possible, the block now has the form of a prefix of extracts followed by a suffix of inserts. The algorithm keeps a stack of hashes of (item, height) pairs, along with the number of items summarized in each hash. Each item extract in the block, along with the current height, is removed from the hash on the top of the stack, until the hash supposedly represents no items. If the hash is not identically zero, the algorithm rejects. Otherwise, the procedure proceeds to the next hash on the stack, until the prefix of extracts are exhausted. Then the inserted items are hashed with their current height, to build a single new hash value which is pushed onto the top of the stack.

**Queue.** Let QUEUE denote the language over interaction sequences that correspond to queues. That is, the state of the queue at any step $j$ can be represented by a string $Q^j$ so that $Q^0 = \emptyset$, $Q^j = Q^{j-1}u$ if $\sigma_j = \text{ins}(u)$ and $Q^j = Q^{j-1}_{2:|Q^{j-1}|}$ if $\sigma_j = \text{ext}(u)$. Then $\sigma \in$ QUEUE for $|\sigma| = N$ iff

$$Q^N = \emptyset \quad \text{and} \quad \forall j \in [N], (\sigma_j = \text{ext}(u) \implies u = Q^{j-1}_1)$$

As observed in [5], it is possible to recognize QUEUE with a single pass and $O(\log N)$ space: we use a single fingerprint to check that the value of the $i$th insert equals the value of the $i$th extract for all $i \in [N]$.

**Deque.** Let DEQUE denote the language over interaction sequences that corresponds to double-ended queues. That is, there are now two types of insert and extract operations, one operation for the head and one for the tail. Clearly, since a deque can simulate a stack via operations on the tail only, recognizing DEQUE is at least as hard as recognizing STACK. For the upper bound, it is possible to adapt the algorithm of [15]. Again, each block of $\sqrt{N}$ operations is partitioned into a prefix of extractions (to head and tail) and insertions (to head and tail). Now we maintain a deque of hash values of item, height pairs. Each extract to the head is applied to the hash at the head of the deque of hashes, and each extract to the tail is applied to the hash at the tail of the deque. The same check is applied: any hash which should now summarize no items must be identically zero (otherwise, the algorithm rejects). Inserts to the head are parceled up into a hash which is placed at the head of the deque, and inserts to the tail are placed in a hash at the tail of the deque. Then we accept $\sigma$ if after processing $\sigma$ the algorithm reaches an empty deque and has not rejected at any point. This gives the following theorem.

**Theorem 4.12.** *Every $O(1/\log^2 N)$-error, $p$-pass, $s$-space randomized streaming algorithm to recognize* DEQUE *on length $N$ streams must satisfy $ps = \Omega(\sqrt{N})$. It is possible to recognize* DEQUE *in one pass with $O(\sqrt{N}\log N)$ bits of space with high probability.*

### 4.3 Variations with timestamps

As noted in the introduction, the results of Blum et al. [5] can be viewed as recognizing languages where each $\text{ext}(u)$ is augmented with the timestamp of its matching $\text{ins}(u)$, and is denoted $\text{ext}(u, t)$. These languages are defined as before, but with the additional constraints that each $t \in [N]$ appears at most once across all extracts and

$$\forall j \in [N], (\sigma_j = \text{ext}(v, t) \implies \sigma_t = \text{ins}(v))$$

This defines the variant languages QUEUE-TS, STACK-TS, DEQUE-TS and PQ-TS. The observations of Blum et al. imply that verifying strings in STACK-TS and QUEUE-TS (and ensuring that all the timestamps are also consistent) requires only $O(\log N)$ space. The same argument also gives an $O(\log N)$ bound for deques. For PQ-TS, the problem seems harder: Chu et al. [8] gave an $\widetilde{O}(\sqrt{N})$ streaming algorithm which relied heavily on the presence of timestamps (and hence does not recognize PQ without timestamps). We leave as an open question the problem of fully resolving the complexity of recognizing priority queue sequences with timestamps, since the reduction via augmented indexing no longer holds in this case.

## A  Proof of the Direct Sum Theorem

For completeness, we give a full proof of our direct sum theorem that relates the information complexity of MULTI-AI with that of AI. We begin with a small technical lemma that is an interesting observation in its own right.

**Lemma A.1.** *Let $P$ be a communication protocol involving two players, Alice and Bob, who share a public random string $R$ in addition to their private random strings. Let $T$ denote the transcript of $P$ when Alice receives input $X$ and Bob receives $Y$, from an arbitrary input distribution. Let $A$ and $B$ denote the portions of $T$ that are communicated by Alice and Bob, respectively. Then*

$$\mathrm{I}(T : X \mid Y, R) \;=\; \mathrm{I}(A : X \mid Y, R) \text{ and } \mathrm{I}(T : Y \mid X, R) \;=\; \mathrm{I}(B : Y \mid X, R).$$

*Proof.* By the chain rule for mutual information, we have

$$\mathrm{I}(T : X \mid Y, R) \;=\; \mathrm{I}(AB : X \mid Y, R) \;=\; \mathrm{I}(A : X \mid Y, R) + \mathrm{I}(B : X \mid A, Y, R).$$

Since Bob's messages are just some function of $A, Y, R$, and his private coins, for any fixed setting of $A, Y, R$, we have that $B$ and $X$ are independent. Thus, $\mathrm{I}(B : X \mid A, Y, R) = 0$. Similarly, we can show that $\mathrm{I}(T : Y \mid X, R) = \mathrm{I}(B : Y \mid X, R)$. □

**Theorem 3.2 (restated).** *Suppose there exists a $[p, s, \varepsilon]$-protocol $Q$ for MULTI-AI$_{m,n}$. Then there exists an $\varepsilon$-error randomized protocol $P$ for AI$_n$ in which Alice sends at most $ps$ bits in total, and which satisfies*

$$m \cdot \mathrm{icost}^B_{\mu_0}(P) \;\leq\; \mathrm{icost}_{\mu_0^{\otimes m}}(Q),$$

*where $\mu_0$ is as in Definition 2.1 and $\mu_0^{\otimes m}$ denotes the $m$-fold product of $\mu_0$ with itself.*

*Proof.* Using $Q$, we can derive a family, $\{P_j\}_{j \in [m]}$, of protocols for AI, using the following simulation. Suppose Alice and Bob receive inputs $x$ and $(k, c, x_{1:k-1})$ respectively.

1. Alice sets $A_j$'s input to $x$ and Bob sets $B_j$'s input to $(k, c, x_{1:k-1})$.

2. The players generate $X^1, X^2, \ldots, X^{j-1}, X^{j+1}, \ldots, X^m, K^1, \ldots, K^{j-1}$ independently and uniformly at random using *public* coins. They choose $C^1, \ldots, C^{j-1}$ so that $X^i_{K^i} = C^i$ for all $i \in [j-1]$. This sets the input to players $A_1, B_1, \ldots, A_{j-1}, B_{j-1}$ and ensures that $(X^i, K^i, C^i) \sim \mu_0$ for all $i < j$.

3. Bob generates $K^{j+1}, K^{j+2}, \ldots, K^m$ independently and uniformly at random using *private* coins. He chooses $C^{j+1}, \ldots, C^m$ so that $X^i_{K^i} = C^i$ for each $i \in \{j+1, \ldots, m\}$. This sets the input to players $A_{j+1}, B_{j+1}, \ldots, A_m, B_m$ and ensures that $(X^i, K^i, C^i) \sim \mu_0$ for all $i > j$.

4. The players now jointly simulate $Q$ on the random input **Z** thus generated. In each round:

    (a) Alice simulates players $A_1, B_1, \ldots, A_j$ and sends Bob the message that $A_j$ would have sent to $B_j$.

    (b) Bob simulates $B_j, A_{j+1}, \ldots, B_m$ and then sends Alice the message that $B_m$ would have sent to $A_m$.

(c) Alice then continues the simulation of $A_m, \ldots, A_1$ and moves on to beginning of the next round (if required), without having to communicate anything.

5. At the end of the simulation, Alice outputs the answer that player $A_1$ would have output in $Q$.

Clearly, Alice communicates at most $ps$ bits in $P_j$. The definition of $\mu_0$ ensures that $\text{AI}(X^i, K^i, C^i) = 0$ for all $i \neq j$, and therefore $\text{MULTI-AI}(\mathbf{Z}) = \text{AI}(X, K, C)$; thus $P_j$ is correct whenever $Q$ is correct on the randomly generated input. This bounds the worst-case error of $P_j$ by $\varepsilon$. To bound the information cost of $P_j$, notice that when the input to $P_j$ is distributed according to $\mu_0$, it simulates $Q$ on an input that is distributed according to $\mu_0^{\otimes m}$. Let $(X^j, K^j, C^j)$ denote a random input to $P_j$ distributed according to $\mu_0$, and let $T$ and $B$ denote the resulting random transcript of $P_j$, and Bob's portion of this transcript, respectively. Defining $M_m$ and $R$ as in (22), we see that $B \equiv M_m$ and that the public random string used by $P_j$ is exactly $R' = (R, \mathbf{X}^{-\mathbf{j}}, K^1, \ldots, K^{j-1})$. Thus,

$$\begin{aligned}
\text{icost}^B_{\mu_0}(P_j) &= I(T : K^j, C^j \mid X^j, R') \\
&= I(B : K^j, C^j \mid X^j, R') \\
&= I(M_m : K^j, C^j \mid K^1, \ldots, K^{j-1}, X^1, \ldots, X^m, R),
\end{aligned}$$

where the second equality follows from Lemma A.1. By the chain rule for mutual information, we have

$$\begin{aligned}
\text{icost}_{\mu_0^{\otimes m}}(Q) &= I(M_m : K^1, C^1, \ldots, K^m, C^m \mid X^1, \ldots, X^m, R) \\
&= \sum_{j=1}^m I(M_m : K^j, C^j \mid K^1, C^1, \ldots, K^{j-1}, C^{j-1}, X^1, \ldots, X^m, R) \\
&= \sum_{j=1}^m I(M_m : K^j, C^j \mid K^1, \ldots, K^{j-1}, X^1, \ldots, X^m, R) \qquad (24) \\
&= \sum_{j=1}^m \text{icost}^B_{\mu_0}(P_j),
\end{aligned}$$

where (24) holds because $X^j$ and $K^j$ completely determine $C^j$, according to the distribution $\mu_0$. Picking $j$ to minimize $\text{icost}^B_{\mu_0}(P_j)$ now gives us $m \cdot \text{icost}^B_{\mu_0}(P_j) \leq \text{icost}_{\mu_0^{\otimes m}}(Q)$. □

## B  Reduction from DYCK(2) to PQ

**Lemma 4.3 (restated).** *There exists an $O(\log N)$-space stream reduction from DYCK(2) to PQ(4N).*

*Proof.* Consider a string $p$ over parentheses $\{a, \bar{a}, b, \bar{b}\}$ and define

$$\text{height}(p) := |\{j : p_j \in \{a, b\}\}| - |\{j : p_j \in \{\bar{a}, \bar{b}\}\}|$$

and $\text{height}(\epsilon) = 0$. Define the transformation $\psi$ by $\psi(p) = \phi(p_{1:1})\phi(p_{1:2})\ldots\phi(p_{1:N})$ where:

$$\phi(p_{1:i}) = \begin{cases} \text{ins}(2N - 2\,\text{height}(p_{1:i-1})) & \text{if } p_i = a \\ \text{ext}(2N - 2\,\text{height}(p_{1:i})) & \text{if } p_i = \bar{a} \\ \text{ins}(2N - 2\,\text{height}(p_{1:i-1}) - 1) & \text{if } p_i = b \\ \text{ext}(2N - 2\,\text{height}(p_{1:i}) - 1) & \text{if } p_i = \bar{b} \end{cases}$$

First, note that the transformation can be done in $O(\log N)$ space since it is sufficient to maintain the height of the last two elements. The transformation is onto $[4N]$, since for any arbitrary string $p$ of $N$ parentheses $-N < \text{height}(p_{1:N-1}) < N$.

We now argue that $p \in \text{DYCK}(2)$ iff $\psi(p) \in \text{PQ}$. For notational convenience, we first define $\psi(p|p') = \phi(p'p_{1:1})\phi(p'p_{1:2})\ldots\phi(p'p_{1:|p|})$ and note that $\psi(p'p) = \psi(p')\psi(p|p')$.

- $p \in \text{DYCK}(2)$ implies $\psi(p) \in \text{PQ}$: We prove this by induction on the length of $p$. We may decompose $p = p^1 c\bar{c} p^2$ where $c \in \{a,b\}$ and $h = \text{height}(p^1)$ is maximal over all such decompositions. Without loss of generality assume $c = a$. Note that $p^1 p^2 \in \text{DYCK}(2)$ and hence, by induction $\psi(p^1 p^2) = \psi(p^1)\psi(p^2|p^1) \in \text{PQ}$. But observe that

$$\psi(p) = \psi(p^1)\psi(b\bar{b}|p^1)\psi(p^2|p^1 b\bar{b}) = \psi(p^1)\psi(b\bar{b}|p^1)\psi(p^2|p^1)$$

  which is in PQ because $\psi(p^1)\psi(p^2|p^1) \in \text{PQ}$ and $\psi(b\bar{b}|p^1) = \text{ins}(2N - 2h - 1)\text{ext}(2N - 2h - 1)$ where $2N - 2h - 1 \leq \{u : \text{ins}(u) \in \psi(p^1)\}$. Since $h$ is maximal, $2N - h - 1$ is indeed the smallest value when it is extracted.

- $p \notin \text{DYCK}(2)$ implies $\psi(p) \notin \text{PQ}$. Since $p \notin \text{DYCK}(2)$, a standard characterization of the language implies that one of the following cases is true:

  - Case 1. $\text{height}(p_{1:N}) \neq 0$. Therefore, there are different numbers of extracts and inserts in $\psi(p)$ and hence $\psi(p) \notin \text{PQ}$, since each open parenthesis maps onto an insert and each close parenthesis maps onto an extract.
  - Case 2. $\text{height}(p_{1:i}) < 0$ for some $i \in [N]$. Therefore, there are more extracts than inserts in a prefix of $\psi(p)$ and hence $\psi(p) \notin \text{PQ}$.
  - Case 3. There exists a smallest $j$ such that for some $i < j$,
    * $\text{height}(p_{1:i-1}) = \text{height}(p_{1:j}) =: h$
    * $p_{1:j-1}$ is a prefix for a string in $\text{DYCK}(2)$ and hence $\psi(p_{1:j-1})$ is a prefix for a string in PQ
    * $(p_i, p_j) = (a, \bar{b})$ or $(p_i, p_j) = (b, \bar{a})$.

  Since $\psi(p_{1:j-1})$ is a prefix for a string in PQ, we can consider the state, $M_{j-1}$, of the priority queue after the interaction sequence $\psi(p_{1:j-1})$ as defined in Definition 4.2. Note that $M_{j-1}$ contains at most one element from $\{2N - 2k - 1, 2N - 2k\}$ for each $k$ (else $j$ was not the minimal choice). If $(p_i, p_j) = (b, \bar{a})$ then $2N - 2h - 1 \in M_{j-1}$. But $\phi(p_{1:j}) = \text{ext}(2N - 2h)$ and we therefore deduce that $\psi(p) \notin \text{PQ}$. If $(p_i, p_j) = (a, \bar{b})$ then $2N - 2h \in M_{j-1}$ and hence $2N - 2h - 1 \notin M_{j-1}$. Since $\phi(p_{1:j}) = \text{ext}(2N - 2h - 1)$, we deduce that $\psi(p) \notin \text{PQ}$.

□